\documentstyle[11pt,aaspp4]{article}

\input tcilatex.tex
\input colordvi.sty 

\received{2002 January 25}
\accepted{2002 May 1}
\begin{document}

\baselineskip 12pt
\parskip 4pt

\title{Dynamics and Structure of Three-Dimensional Trans-Alfv\'enic Jets. II. The Effect of Density and Winds}

\author{Philip E. Hardee}

\affil{Department of Physics \& Astronomy \\The University of Alabama \\Tuscaloosa,
AL 35487 \\hardee@athena.astr.ua.edu}

\author{Alexander Rosen}
 
\affil{Armagh Observatory \\ College Hill \\ Armagh, BT61 9DG, Northern Ireland
 \\rar@star.arm.ac.uk}

\begin{abstract}
\footnotesize
\baselineskip 10pt

Two three-dimensional magnetohydrodynamical simulations of strongly
magnetized conical jets, one with a poloidal and one with a helical
magnetic field, have been performed. In both simulations the jet is
precessed at the origin to break the symmetry. The jets are denser by
an order of magnitude than the jets in a previous set of simulations.
Theoretical work accompanying the previous simulations indicated that
an increase in jet density would stabilize the jets beyond the Alfv\'en
point and this prediction is verified by the present simulations.  In
the poloidal simulation a significant sheath of magnetized moving
material developed around the jet and led to additional stabilization.
New theoretical work analyzing the effect of a jet embedded in a
magnetized wind shows that the velocity shear, $\Delta u \equiv u_j -
u_e$ must exceed a ``surface'' Alfv\'en speed, $V_{As} \equiv [(\rho
_j+\rho _e)(B_j^2+B_e^2)/(4\pi \rho _j\rho _e)]^{1/2}$, before the jet
becomes unstable to helical and higher order modes of jet distortion
and this explains the enhanced stability found in the poloidal
simulation.  The fundamental pinch mode is not similarly affected and
emission knots with spacing $\sim 4R_j$ developed in the poloidal
simulation.  Thus, we identify a mechanism that can suppress large
scale asymmetric structures while allowing axisymmetric structures to
develop, and astrophysical jets surrounded by outflowing winds will be
more stable than if surrounded by a stationary or backflowing external
medium. Knotty structures along a straight protostellar jet like the
jet in HH\,34 or in the inner part of the jet in HH\,111 could be
triggered by pinching of a low magnetosonic Mach number jet surrounded
by a suitable wind.  Of additional interest is the development of
magnetic field orientation along the line-of-sight organized by the
toroidal flow field accompanying helical twisting.  On astrophysical
jets such structure could lead to a reversal of the direction of
Faraday rotation in adjacent zones along a jet.  Thus, Faraday rotation
structure like that seen along the 3C\,465 jet could be attributed to
organized magneto-hydrodynamical structures produced by the jet flow.

\end{abstract}

\keywords{galaxies: jets --- hydrodynamics --- instabilities --- MHD --- ISM: jets and outflows}

\baselineskip 12pt
\parskip 2pt

\vspace{-1.0 cm}
\section{Introduction}

Jet acceleration and collimation schemes imply dynamically strong
magnetic fields close to the central engine (see Meier, Koide, \&
Uchida 2001).  Numerical studies, e.g., Meier, Payne, \& Lind (1996),
Ouyed, Pudritz, \& Stone (1997), Ouyed \& Pudritz (1997), Romanova et
al.\ (1997), show that the jets created in this fashion pass through
slow magnetosonic, Alfv\'enic and fast magnetosonic critical points,
whose ultimate velocity may depend on the configuration of the magnetic
field (Meier et al.\ 1997) and accelerate up to speeds that may be only
a few times the Alfv\'en velocity at the Alfv\'en surface (Camenzind
1997).  Thus, low magnetosonic Mach numbers might be expected for
magnetically launched jets (Fendt \& Camenzind 1996; Lery \& Frank
2000).  The numerical studies and also theoretical studies (Begelman \&
Blandford 1999; Lery et al.\ 2002) indicate that a wind or magnetized
wind may surround a more rapidly moving jet.

All jet flows are susceptible to Kelvin-Helmholtz (KH) or current
driven (CD) instabilities. The KH instability of 3D jets with purely
poloidal or purely toroidal magnetic fields (Ray 1981; Ferrari,
Trussoni, \& Zaninetti 1981; Fiedler \& Jones 1984; Bodo et al.\ 1989)
and of jets containing force-free helical magnetic fields (Appl \&
Camenzind 1992; Appl 1996) has been extensively investigated for a
stationary external medium.  Additional investigations have considered
the role of CD pinching of toroidally magnetized plasma columns
(Begelman 1998). The CD instabilities are sensitive to the magnetic
profile (Appl, Lery, \& Baty 2000) and, at least for jets containing
force-free helical magnetic fields, it appears that the KH instability
exhibits faster growth and is more likely to be responsible for
producing asymmetric structure than CD instability (Appl 1996) and CD
modes appear confined to the jet interior (Appl, Lery, \& Baty 2000;
Lery, Baty, \& Appl 2000).  In general, KH growth rates increase as the
magnetosonic Mach number decreases provided the jet is
super-Alfv\'enic, but the magnetized jet is nearly completely
stabilized to the KH instability when the jet is sub-Alfv\'enic (Hardee
\& Rosen 1999, hereafter Paper I). In Paper I the strongly magnetized
``light'' jets experienced considerable mass entrainment and slowing as
denser material was entrained following destabilization.  Denser jets
are predicted to remain stable beyond the Alfv\'en point and, in
general, have been found to be more robust than their less dense
counterparts (Hardee, Clarke, \& Rosen 1997).

In this paper we analyze results from two simulations, one with a
poloidally magnetized jet and the other with a helically magnetized
jet.  The jets are an order of magnitude more dense than those studied
in Paper I. In \S 2 the numerical setup and results of the numerical
simulations are presented.  The simulations are initialized by
establishing a cylindrical helically magnetized jet across a
computational grid in an unmagnetized surrounding medium with pressure
gradient devised to result in an approximate constant expansion of the
jet once pressure equilibrium is achieved, i.e., produce a conical
jet.  In one simulation a significant outwards flowing magnetized
cocoon develops that influences jet stability and dynamics. This
simulation shows how a jet might be influenced by a surrounding
magnetized wind.  Synchrotron intensity images and an integration of
the line-of-sight magnetic field provide a connection between jet
dynamics and observable jet emission and polarization structures. In \S
3 the theory is expanded to include a moving magnetized wind medium,
and in \S 4 the predicted effect of density and winds is compared to
the simulation results.  Finally, in \S 5 we discuss some of the
implications for astrophysical jets.

\vspace{-0.7cm}
\section{Numerical Simulations}

\vspace{-0.2cm}
\subsection{Initialization}

Simulations were performed using the three dimensional MHD code
ZEUS-3D, an Eulerian finite-difference code using the Consistent Method
of Characteristics (CMoC) which solves the transverse momentum
transport and magnetic induction equations simultaneously and in a
planar split fashion (Clarke 1996). Interpolations were carried out by
a second-order accurate monotonic upwinded time-centered scheme (van
Leer 1977) and a von-Neumann Richtmyer artificial viscosity was used to
stabilize shocks. The code has been thoroughly tested via MHD test
suites as described by Stone et al.\ (1992) and Clarke (1996) to
establish the reliability of the techniques.

All simulations are initialized by establishing a cylindrical jet
across a 3D Cartesian grid resolved into 130 $\times $ 130 $\times $
370 zones.  Thirty uniform zones span the initial jet diameter, $2R_0$,
along the transverse Cartesian axes ($x$-axis and $y$-axis). Outside
the uniform grid zones, the grid zones are ratioed where each
subsequent zone increases in size by a factor 1.05. Altogether the 130
zones along the transverse Cartesian axes span a total distance of
$30R_0$. Along the $z$-axis 280 uniform zones span a distance of
$40R_0$ outwards from the jet origin. An additional 90 ratioed zones
span an additional distance of $40R_0$ where each subsequent zone
increases in size by a factor 1.02. The 370 zones along the $z$-axis
span a total distance of $80R_0$. Outflow boundary conditions are used
except where the jet enters the grid where inflow boundary conditions
are used. The use of a non-uniform grid with larger zones at the grid
boundaries has been shown to have the beneficial effect of reducing
reflections off the grid boundaries as a result of increased
dissipation of disturbances by the larger zone size at the grid
boundaries (Bodo et al.\ 1995).

The jets are initialized with a uniform density $\rho _j$ and initial
radius $R_0$. The magnetic field in the jet is initialized with a
uniform axial component, $B_z$, and a toroidal magnetic component with
functional form $B_\phi =B_\phi ^{pk}\sin ^2[\pi f(r)]$ where for
$r<r_{pk}$, $f(r)=0.5(r/r_{pk})^a$, and for $r_{pk}<r<r_{\max }$, $
f(r)=1.0-0.5\left[ \left( 1-r/r_{\max }\right) /(1-r_{pk}/r_{\max
})\right] ^b$. In these simulations the toroidal component increases to
a maximum, $ B_\phi ^{pk}$, at $r_{pk}=0.5R_0$, and declines to zero
at $r_{\max }=0.9R_0 $ so that initially all currents flow within the
jet. In the simulation with primarily axial field (simulation P),
$a=1.106$ and $b=0.885$, and the FWHM of the toroidal field is $0.5R_0$. In
the simulation with helical magnetic field (simulation T), $a=0.346$ and
$b=0.277$, and the FWHM of the toroidal field is $0.8R_0$. These
particular toroidal profiles are not physically motivated but are well
behaved numerically and in simulation T the toroidal profile provides a
broad cross sectional region within the jet in which the toroidal
magnetic component is relatively constant (nearly constant helical
pitch, plasma beta, and magnetosonic speed). In the external medium the
magnetic field is equal to zero. The equation of hydromagnetic
equilibrium
$$
\frac d{dr}\left( p_{jt}(r)+\frac{B_z^2(r)}{8\pi }+\frac{B_\phi ^2(r)}{8\pi }
\right) =-\frac{B_\phi ^2(r)}{4\pi r}\text{ ,} 
$$
where the term on the right hand side describes the effects of magnetic
tension, has been used to establish a suitable radial gas pressure
profile in the jet by varying the jet temperature. The sound, Alfv\'en
and magnetosonic speeds are defined as $a^2\equiv \Gamma p/\rho $ with
$\Gamma =5/3$, $V_A^2\equiv B^2/4\pi \rho $, and $a_{ms}^2\equiv
a^2+V_A^2$. Since internal dynamics and timescales involve wave
propagation across a jet with propagation speed a function of jet
radius we define radial averages as, for example,
$$
\left\langle M_{ms}\right\rangle \equiv \frac 1{R_j}\int_0^{R_j}M_{ms}(r)dr
\text{ ,} 
$$
that will differ for different magnetic, density, and temperature profiles.

In the simulations the external medium is isothermal and the external
density $\rho _e(z)$ declines to produce a pressure gradient,
$p_e(z)\propto \rho _e(z)$, that leads to a constant expansion,
$R_j(z)=(1+z/80R_0)R_0$, of a constant velocity adiabatic jet
containing a uniform poloidal magnetic field and an internal toroidal
magnetic field that provides some confinement, i.e.,
$$
\rho _e(z)=\frac{\left[ (R_j/R_0)^{-10/3}+C_p(R_j/R_0)^{-4}-C_\phi
(R_j/R_0)^{-2}\right] }{\left[ 1+C_p-C_\phi \right] }\rho _e(0) \text{ .}
$$
The values of $C_p$ and $C_\phi $ depend on the poloidal and toroidal
field strengths, and the ratio of the magnetic pressure relative to the
thermal pressure. A pseudo-gravitational potential keeps the external
medium static.  Simulation P contains a primarily axial magnetic
field configuration with only a weak toroidal component, $B_\phi
^{pk}/B_z\sim 0.05$ at the inlet, so as to be directly comparable to
predictions made by a linear stability analysis containing a $B_z$
magnetic field component only (\S 3). In simulation T, the axial
magnetic field strength is identical to simulation P but now the
toroidal component $B_\phi ^{pk}/B_z\sim 0.44$ at the inlet. The jets
are initialized with a density $\rho _j=0.20\rho _e(0)$ and with a
velocity $u_j=1.78a_e$ that is trans-Alfv\'enic. The jet sound,
Alfv\'en and magnetosonic speeds at jet center normalized to the
external sound speed and radial averages of the jet sonic, Alfv\'enic
and magnetosonic Mach numbers at the inlet and values of $C_p$ and
$C_\phi $ for the simulations are given in Table 1. Absolute values for
jet speed, density, temperature, and magnetic field strength are
completely determined by choosing values for the external density,
$\rho _e(0)$, and the external temperature, $T_e$, or sound speed,
$a_e$. An absolute length scale is determined by choosing a value for
$R_0$.
\vspace{-0.75cm}
\begin{table} [h]
 \begin{center}
 \caption{Initial Conditions \label{tbl-1}}
 \vspace{0.1cm}
 \begin{tabular}{ c c c c c c c c c c} \hline \hline
     {\bf Simulation} & {$u_j/a_e$}  &  {$a_{j}/a_{e}$}  &  {$V_{A,j}/a_{e}$}  &  { $a_{j}^{ms}/a_{e}$}   &  {$\left\langle M_{j}\right\rangle$}  &  {$\left\langle M^{A}_{j}\right\rangle$}   &  {$\left\langle M^{ms}_{j}\right\rangle$}  & { $C_p$}  & {$C_{\phi}$}  \\ \hline
  P & 1.78 & 1.32 & 1.90 & 2.31 & 1.43 & 0.89 & 0.75 & 1.721 & 0.005    \\
  T & 1.78 & 1.41 & 1.90 & 2.37 & 1.26 & 0.87 & 0.70 & 1.506 & 0.294    \\
 \hline
\end{tabular}
\end{center}
\end{table}
\vspace{-0.5cm}
In both simulations the jet is driven by a periodic precession of the
jet velocity, $u_j$, at an angle of 0.01 radian relative to the
$z$-axis with an angular frequency $\omega =0.5a_e/R_0$. The initial
transverse motion imparted to the jet by this precession is well within
the linear regime. The precession serves to break the symmetry and the
precessional frequency is chosen to be below the theoretically
predicted maximally unstable frequency associated with helical twisting
of a KH unstable supermagnetosonic jet with jet radius $R_j\geq R_0$.
In both simulations the precession is in a counterclockwise sense when
viewed outwards from the inlet. This direction of precession induces a
helical twist in the same sense as that of the magnetic field helicity.

In both simulations the jets expand rapidly and after about five
dynamical times, $\tau _d\equiv (a_e/R_0)t\approx 5$, have achieved an
approximate static pressure equilibrium with the surrounding medium,
and a nearly constant expansion rate on the computational grid with $
R_j\approx 2R_0$ when $z=80R_0$.
\begin{figure}[hb]
\vspace{9.0cm}
\includegraphics{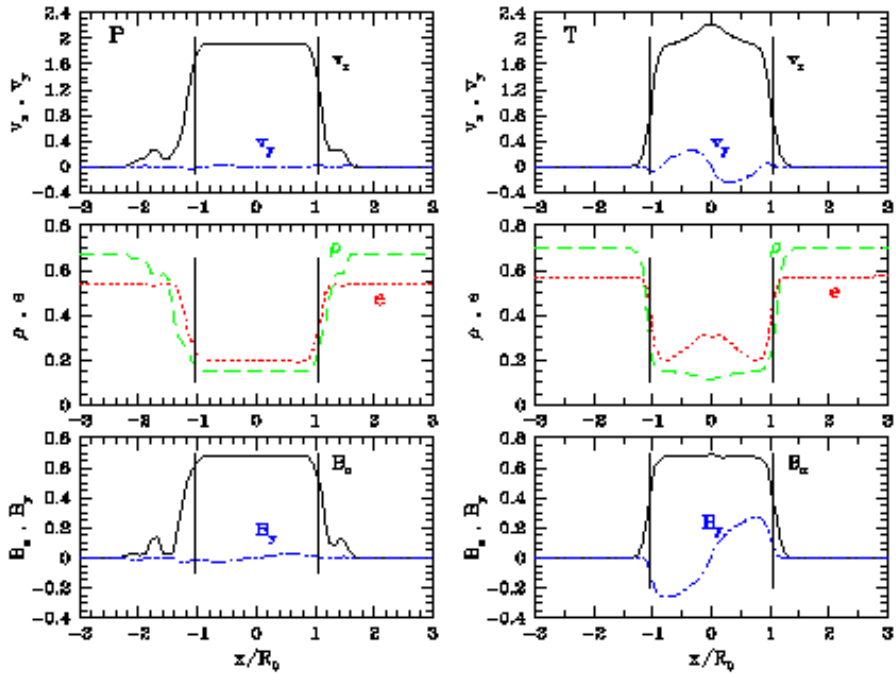}
\caption {\footnotesize \baselineskip 10pt Profiles (along the $x$-axis) of axial and azimuthal velocity, $v_z$ \& $v_y$, respectively, density and internal energy, $\rho$ \& $e$, and magnetic field components, $B_z$ \& $B_y$ for simulations P (left) and T (right) at $z/R_0 = 4.5$. The vertical lines indicate the location of a conical jet`s surface at this distance from the inlet.}
\end{figure}
The code modifies the initial input
static equilibrium state to an appropriate dynamic equilibrium state,
i.e., $\nabla \times ({\bf u}\times {\bf B})=0$ throughout the duration
of the simulation within an axial distance of one jet radius from the
inlet. Velocity, density, internal energy and magnetic field profiles
representative of the initial dynamic equilibrium in the two
simulations are shown in Figure 1.

The simulations were terminated at dynamical times (P) $\tau _d\equiv
(a_e/R_0)t=96$ and (T) $\tau _d\equiv (a_e/R_0)t=56$ when computational
difficulties arose.  At these times the simulations have reached a
quasi-steady state out to axial distances of about (P) $60R_0$ and (T)
$35R_0$. At this time the jet had undergone (P) $\sim 21.4$ and (T)
$\sim 12.5 $ precessional periods and (P) $\sim 3.7$ and (T) $\sim 2.2$
flow through times, $\tau_{fl} \equiv (z/u)(a_e/R_0)$, through
$z=50R_0$.  Typically a quasi-steady state is achieved after 3 flow
through times at a particular distance. Each simulation required
roughly 110 CPU hours on the Cray T90 at the San Diego Supercomputer
Center.

\vspace{-0.5cm}
\subsection{Simulation Results}

\vspace{-0.2cm}
\subsubsection{Jet Structure}

Plots of the total pressure, $p^{*}\equiv p_{th}+p_{bz}$, normalized by
the external pressure p$_e$ and velocity components along 1D cuts
parallel to the jet axis in the $x$-$z$ slice plane are shown in Figure
2. 
\begin{figure}[h]
\vspace{11.75cm}
\includegraphics{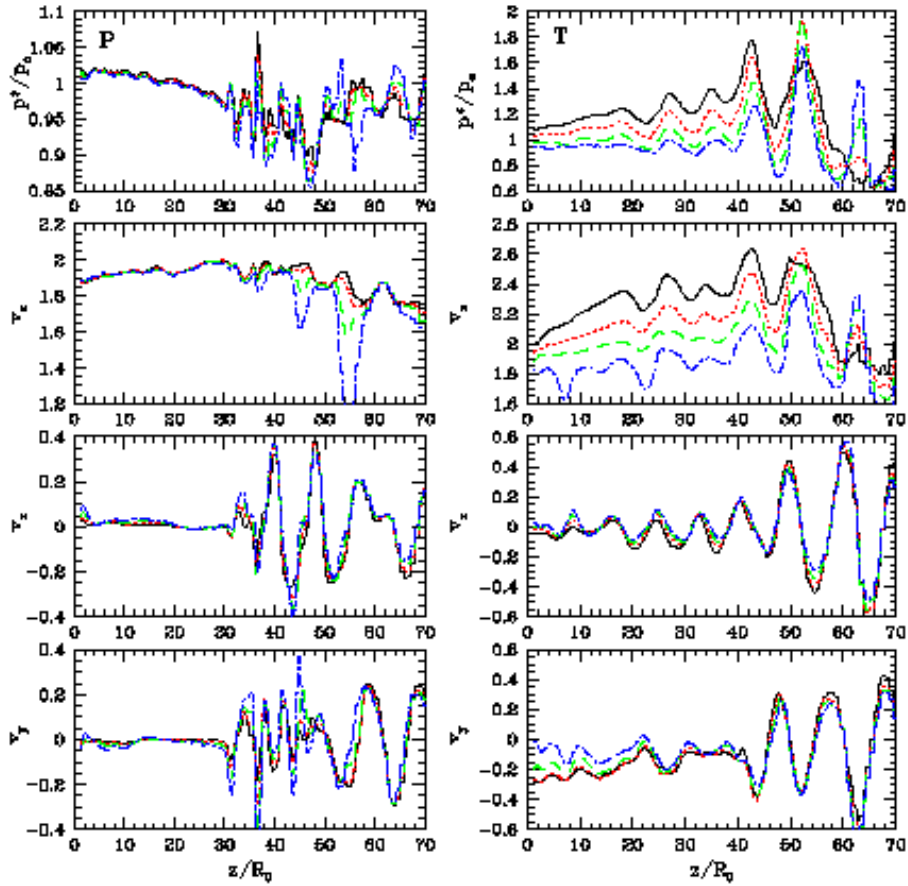}
\caption {\footnotesize \baselineskip 10pt Cuts in the $x$-$z$ plane parallel to the $z$-axis at $x/R_0 =$ (solid line) 0.233, (\Red{dotted line}) 0.433, (\Green{dashed line}) 0.633, (\Blue{dot-dash line}) 0.833 of jet total pressure, $p^{*}$, normalized by the external thermal pressure, $p_e$, and the three velocity components normalized by the external sound speed, $a_e$, for simulations P (left) and T (right).}
\end{figure}
In this slice plane $v_x$ and $v_y$ correspond to a radial velocity,
$v_r$, and an azimuthal velocity, $v_\phi $, respectively, provided the
jet center is not displaced off the $z$-axis. The jet accelerates in
response to thermal and magnetic pressure gradients, and to the
pseudo-gravitational potential. In simulation P with primarily poloidal
magnetic field the jet accelerates to 5\% above the value at the inlet.
In simulation T additional acceleration occurs, to about 20\% above the
inlet value, as a result of the ``spring'' effect from the helical
magnetic field.  Instability is most obviously manifested by the
oscillations and growth in the amplitude  of the transverse velocity
components, but also appears in total pressure and axial velocity
fluctuations. Low level fluctuations in $p^{*}$ and $v_z$ appear in
both simulations by $z\sim 5R_0$.  Significant fluctuations do not
begin to grow until $z\sim 30R_0$ in simulation P, but become
significant by $z\sim 15R_0$ in simulation T.

\vspace{0.1 cm}
\noindent
In simulation P we can identify the following features:

\vspace{-0.3 cm}
\begin{enumerate}
\item  At $z/R_0<30$ a short wavelength, $\lambda \lesssim 5R_0$, low
amplitude oscillation in $p^{*}$ (also in $v_z$ but only one
oscillation obvious with the velocity scale used in the figure).

\vspace{-0.2 cm}
\item At $30<z/R_0<50$ a region of rapid growth and complex structure,
showing rapid oscillations in $v_y$ ($v_\phi$) and a dominant
oscillation in $v_x$ ($v_r$) with $\lambda \lesssim 8R_0$.

\vspace{-0.2 cm}
\item  At $z>55R_0$ dominant out of phase transverse velocity
oscillations indicative of helical twisting with $\lambda \sim 10R_0$
moving with $v_w\sim 0.75 a_e \sim 0.40u_j$ where $u_j \sim 1.9 a_e$
(v$_w$ determined from sequential simulation frames).
\end{enumerate}

\vspace{-0.2 cm}
\noindent
In simulation T we can identify the following features:

\vspace{-0.3 cm}
\begin{enumerate}
\item At $z<40R_0$ a region with growing amplitude oscillations in
$p^{*}$, $v_z$ and $v_x$ with $\lambda \lesssim 8R_0$.  

\vspace{-0.2 cm}
\item  At $z>40R_0$ dominant out of phase transverse velocity
oscillations indicative of helical twisting with $\lambda \sim 10R_0$
moving with $v_w\sim 0.91a_e \sim 0.41u_j$ where $u_j \sim 2.2 a_e$.
\end{enumerate}

Plots of the axial velocity along with the sonic, Alfv\'enic and
magnetosonic speeds along 1D cuts parallel to the jet axis in the
$x$-$z$ slice plane inside and outside the jet are shown in Figure 3.
\begin{figure}[h!]
\vspace{7.7cm}
\includegraphics{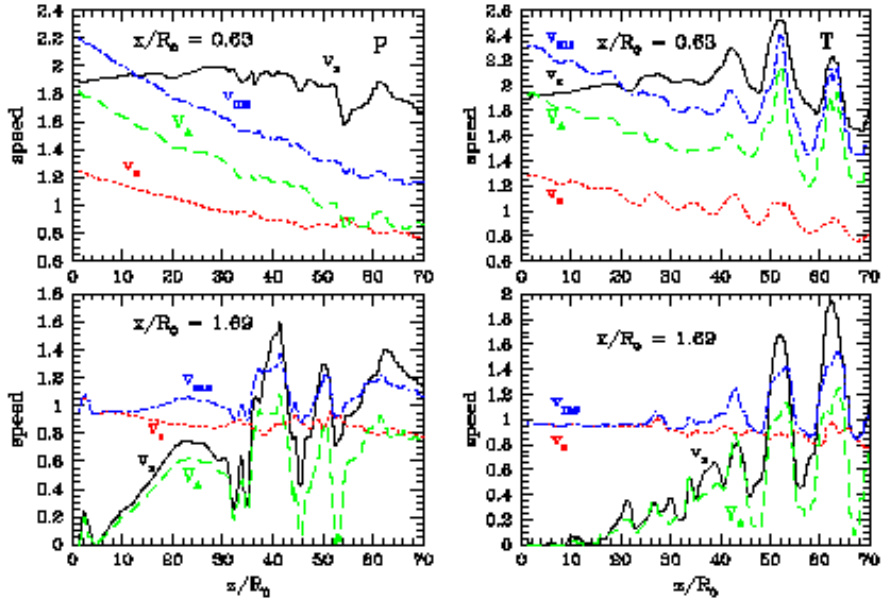}
\caption {\footnotesize \baselineskip 10pt Cuts in the $x-z$ plane parallel to the
$z$-axis inside the jet (top) and outside the jet (bottom) in
simulations P (left) and T (right) showing the (solid line) axial flow speed, $v_z$,
the (\Blue{dot-dash line}) magnetosonic speed, $v_{ms}$, the (\Green{dashed line})
Alfv\'en speed, $V_A$, and the (\Red{dotted line}) sound speed, $v_s$.}
\end{figure}

\noindent
The cut at $x/R_0=1.69$ contacts the conical jet's ``surface'' at
$z=55R_0$ but adequately represents conditions immediately outside the
jet at $z<40R_0$. The cut inside the jet shows that, in both
simulations, the jets are super-Alfv\'enic just outside the inlet and are
super-magnetosonic by $z=20R_0$. The cut outside the jet reveals
that a significant boundary layer forms outside the jet in simulation P
by $z=15R_0$ with flow and Alfv\'en speeds almost half their
values inside the jet.

The differences in transverse structure in the
simulations is shown by velocity, density, internal energy and
magnetic field profiles at $z=18R_0$ shown in Figure 4. 
In both simulations the boundary layer has a thickness greater than
$0.5R_0$ but only in simulation P is there significant flow and
magnetic field in the boundary layer.

\begin{figure}[h]
\vspace{8.7cm}
\includegraphics{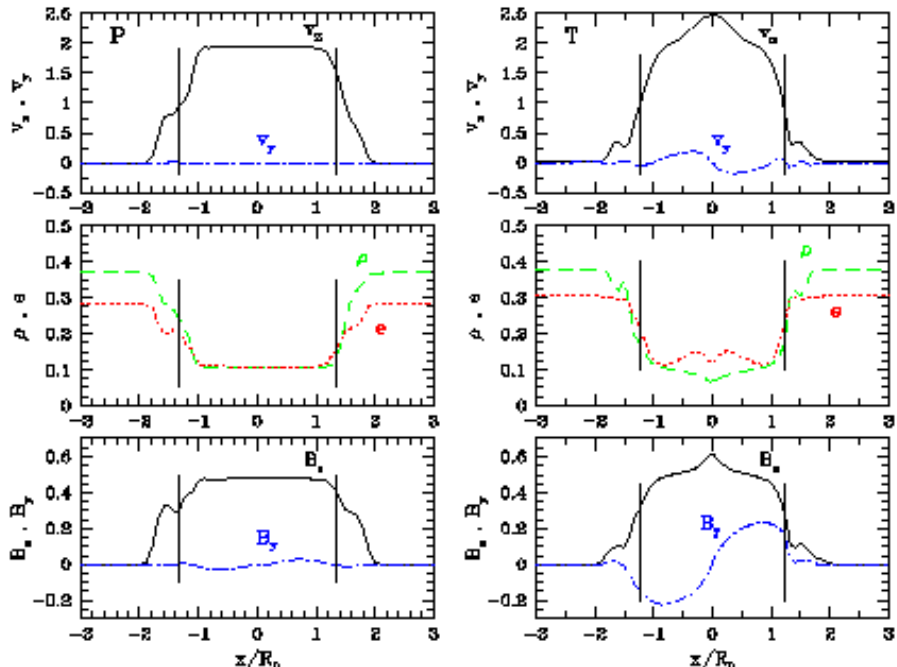}
\caption {\footnotesize \baselineskip 10pt Profiles (along the $x$-axis) of axial and azimuthal velocity, $v_z$ \& $v_y$, respectively, density and internal energy, $\rho$ \& $e$, and magnetic field components, $B_z$ \& $B_y$ for simulations P (left) and T (right)at $z/R_0 = 18.0$. The vertical lines indicate the location of a conical jet`s surface at this distance from the inlet.}
\end{figure}

\vspace{-0.8cm}
\subsubsection{Mass Entrainment}

The mass entrained by the jets and the average velocity of jet plus
entrained material in the two simulations is plotted in Figure 5. In
particular, we define the mass per unit length, $\sigma (z)$, at any
point along the jet as $\sigma =\int_Af\rho dydx$, where A is the cross
sectional area of the computational domain at axial position $z$, and
$f$ is set to 1 if the local magnetic field is above 4\% of the
expected maximum field strength along the jet at $z$ [$B(z)
>0.04B_{j,max}(z)$] and $f$ is set to 0 otherwise.  We define the
entrained mass by the presence of a magnetic field since only the jet
material is initially magnetized. The setting of $f$\ to 1 or 0
effectively assumes that zones with a fraction of jet material, as
defined by the presence of magnetic field, are considered mixed with
the external medium in that zone.  While one may expect that
flux-freezing will prevent the magnetized mass from increasing along
the grid, here the jets are unstable.  Vorticies at the jet's surface
lead to intertwining of filaments or sheets of jet and external
material, and numerical diffusion leads to mixing.  This technique
provides results similar to estimating mass entrainment by using an
axial velocity threshold (Rosen, Hardee, Clarke, \& Johnson 1999;
hereafter RHCJ).  Setting the switch at a magnetic field strength of
4\% of the expected maximum strength at $z$ reduces the sensitivity of
the value of $\sigma $ to a small numerical diffusion of the field into
the denser ``unmixed'' external medium.

\begin{figure}[htb]
\vspace{7.5cm}
\includegraphics{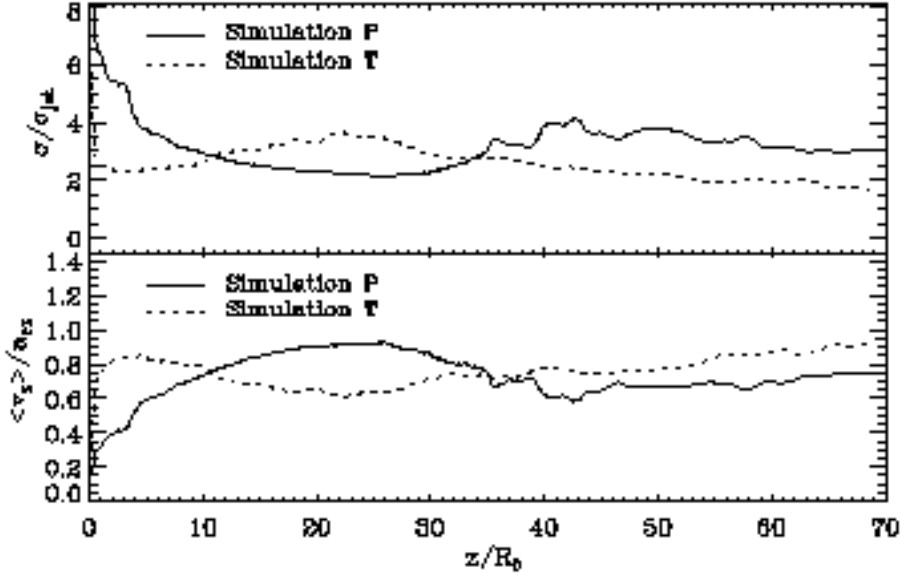}
\caption {\footnotesize \baselineskip 10pt Profiles of the total magnetized mass (jet plus entrained material) per unit length, $\sigma (z)$, and of the average speed of magnetized jet plus entrained material.}
\end{figure}

In the very light jet simulations presented in paper I, there was a
large increase in $\sigma$ near the jet inlet and $\sigma /\sigma _{jet} =
10$ ($\sigma _{jet}$ is the expected value at the inlet) was used as a
baseline value. In these denser jet simulations some immediate increase
is observed but now $\sigma /\sigma _{jet} \sim 2.5$ serves as an
appropriate baseline value.  In simulation P an increase in $\sigma
/\sigma _{jet}$ and a decrease in $\langle v_z\rangle /a_{e}$ at $z \sim
30R_0$ reflects the point at which significant velocity and pressure
fluctuations appear. Simulation T shows a small increase in $\sigma
/\sigma _{jet}$ and a decrease in $\langle v_z\rangle /a_{e}$ at $z\sim
10R_0$ coincident with growth of significant velocity and pressure
fluctuations.

The maximum value of the mass per unit length is $(\sigma /\sigma
_{jet})^{\rm max} \approx$ (P) 4 and (T) 3.5.  These values imply a jet
plus entrained mass of about (P) 1.6 and (T) 1.4 times the initial
baseline mass per unit length.  The mass entrained in simulation P is
similar to the poloidally magnetized simulation B in paper I.  The mass
entrained in simulation T is roughly half that in the toroidal
simulations C \& D in paper I, but this difference is attributable to
the early termination of simulation T before a quasi-steady state could
be achieved at large $z$, e.g., note the decline in the jet plus
entrained mass at larger axial distances.  While the effect of mass
entrainment on the jets here is somewhat reduced relative to the
``light'' jets studied in paper I, we again find that significant mass
entrainment does not occur until the jets destabilize.

The average axial velocity of magnetized material indicates that the
entrained material is moving more rapidly than was found in the light
jet simulations in paper I.  This result indicates that a denser jet is
capable of accelerating entrained material to higher velocities.  In
simulation P we note that the average axial velocity of magnetized
material increases significantly to a maximum at $z/R_0 \sim 25$ while
at the same time $\sigma/\sigma_{jet}$ decreases to a minimum.  This
behavior is coincident with the formation of a significant sheath of
magnetized material moving at up to 38\% of the jet speed (see Figure
3 for velocities in the sheath).  In both simulations the majority of
the momentum flux, $\sigma v_z^2$, [(P) 98\% and (T) 92\%] evaluated
within the quasi-steady state region of the computational grid is
carried by jet plus entrained material, i.e., material with $B(z)
>0.04B_{j,max}(z)$.

\vspace{-0.3cm}
\subsubsection{Velocity, Emission \& Magnetic Field Images}

Jet axial velocity cross sections shown in Figure 6 illustrate
development of the surface distortions that promote mixing and mass
entrainment, and the large scale distortions that can move the jet flow
completely off the initial axis.
\begin{figure}[h]
\vspace{11.3cm}
\includegraphics{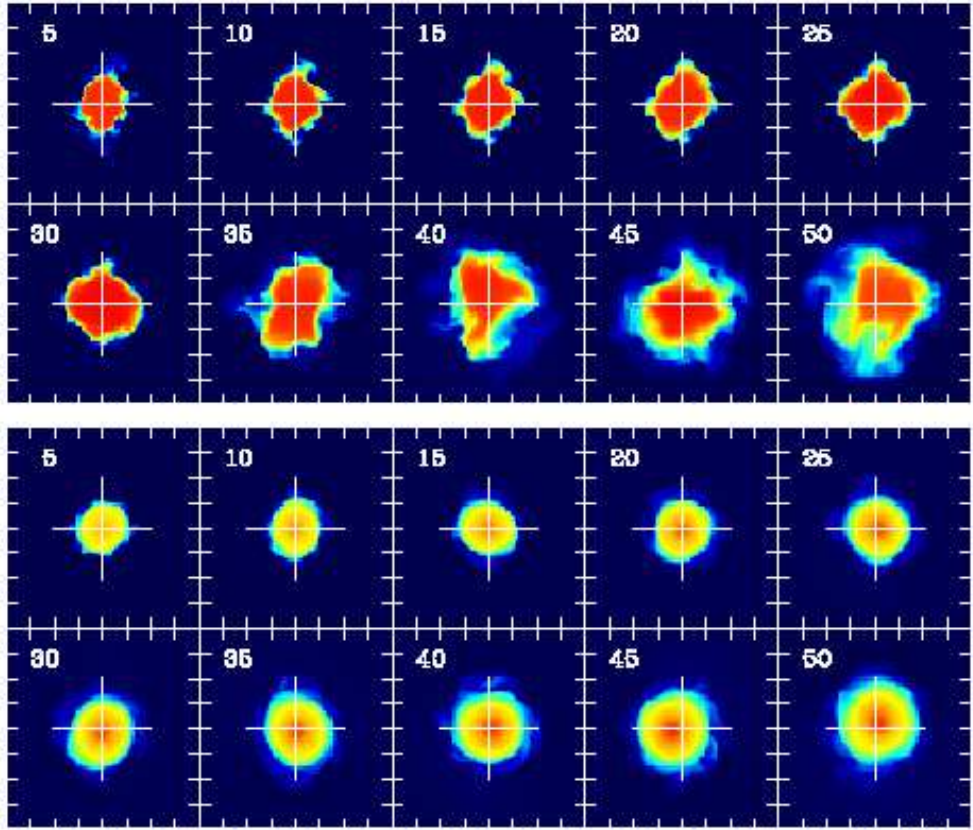}
\caption {\footnotesize \baselineskip 10pt Jet velocity cross sections at axial distances indicated in the panels for simulation P (top 10 panels) and simulation T (bottom ten panels).  The number in the upper left corner of each panel indicates distance from the inlet in units of $R_0$.}
\end{figure}
In simulation P small scale surface corrugations form before
the jet becomes grossly unstable.  A similar result was obtained in
paper I for ``light'' poloidally magnetized jets. An elliptical
distortion of the jet cross section is evident in the panels at $z/R_0=
30$ \& $35$, and considerable cross section distortions and
displacement of the jet off the initial axis occur at larger $z$.
In simulation T the jet exhibits much reduced distortion and the higher
order surface corrugations that appear in simulation P at $z<35R_0$ are
suppressed. These results are similar to what was found in paper I. An
elliptical distortion is evident in the panels $z/R_0 = 10 - 30$ and
the jet is displaced off the initial axis for $z>25R_0$.

Synchrotron intensity images and images of an integration of the
magnetic field component, $B_y$, along the line-of-sight are  shown in
Figures 7 \& 8. To some extent these images also reveal the extent of
jet spreading as only the jet fluid is magnetized.  To generate the
synchrotron intensity images a synchrotron emissivity is defined by
$p_j(B\sin \theta )^{3/2}$ where $\theta $ is the angle made by the
magnetic field with respect to the line of sight. This emissivity
mimics synchrotron emission from a system in which the energy and
number densities of the relativistic particles are proportional to the
energy and number densities of the thermal fluid.  This simplistic
assumption is necessary when the relativistic particles are not
explicitly tracked (Clarke, Norman, \& Burns 1989).

\begin{figure}[ht!]
\vspace{6.5cm}
\includegraphics{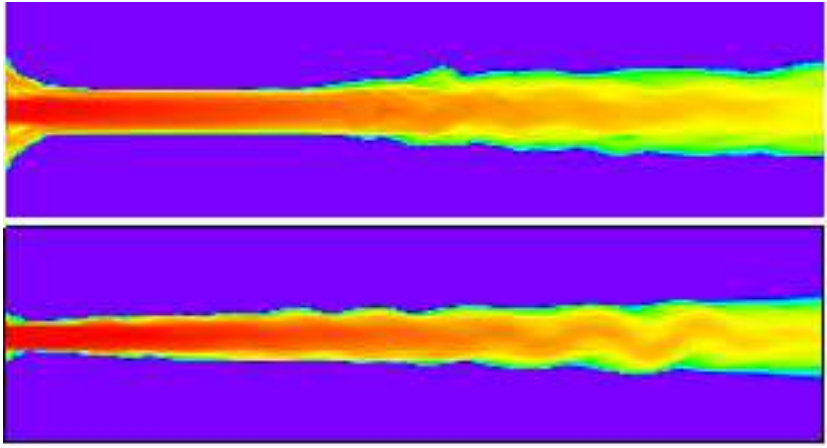}
\caption {\footnotesize \baselineskip 10pt Synchrotron intensity images in the $x$-$z$ plane from
line-of-sight integrations through the computational domain along the $y$-axis.  Image dimensions are $20R_0
\times 80R_0$ and the logarithmic scale spans six orders of magnitude.
Simulations P and T are top and bottom images, respectively.}
\end{figure}

\begin{figure}[htb!]
\vspace{6.4cm}
\includegraphics{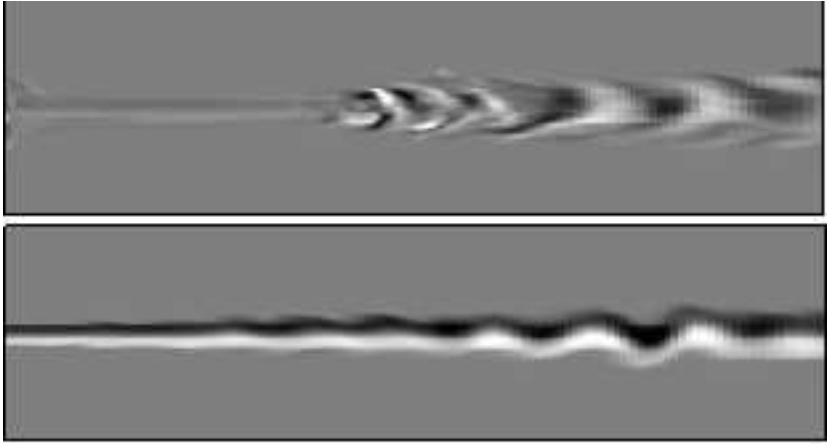}
\caption {\footnotesize \baselineskip 10pt Integration showing the line-of-sight
direction of the magnetic field ($B_y$) in simulations P (top) and T
(bottom).  Image dimensions are identical to Figure 7.}
\end{figure}

The line-of-sight magnetic field changes direction in adjacent bands in
simulation P.  The effect is most pronounced at larger $z$ where a
pronounced helical twist has developed in the jet flow.  The poloidal
magnetic field is organized by the toroidal flow field accompanying the
helical twist.  The line-of-sight magnetic image for simulation T
reflects the orientation of the initial toroidal field component. In
simulation T the toroidal magnetic field is too strong to be overcome
by reorientation of the poloidal field by the toroidal flow field
accompanying the helical twist.

\vspace{-0.3cm}
\subsubsection{Pinch Structure at Intermediate Time}

At dynamical times $\tau_d \sim 39$ the simulations have not yet
developed strong asymmetries.  In Figure 9 synchrotron intensity
images, containing fractional polarization {\bf B}-vectors formed from
the Stokes parameters, reveal observational structures that develop on
the trans-Alfv\'enic jets at intermediate times.  
\begin{figure}[h]
\vspace{6.5cm}
\includegraphics{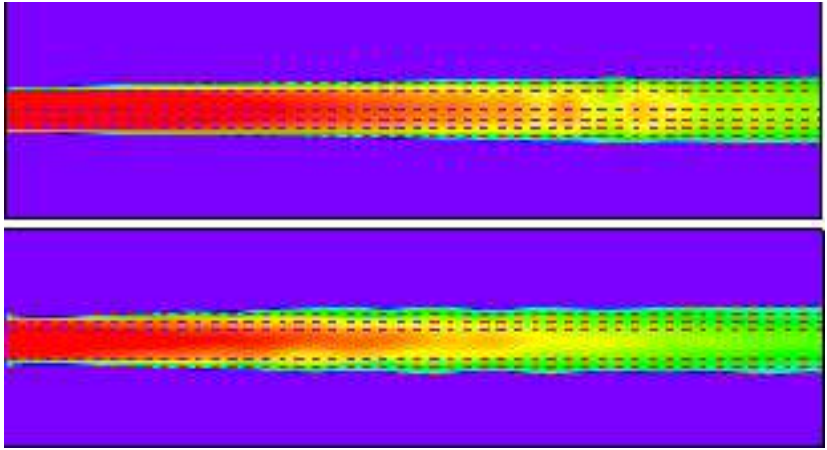}
\caption {\footnotesize \baselineskip 10pt Synchrotron intensity images in the $x$-$z$ plane from
line-of-sight integrations through the computational domain along the
$y$-axis.  Vectors indicate apparent magnetic field direction. Image
dimensions are $10R_0 \times 40R_0$ and simulations P and T are top and
bottom images, respectively.}
\end{figure}
Axisymmetric intensity knots appear contained within the jet interior
in simulation P and are closely spaced.  Intensity knots do not appear
in simulation T and the development of sinusoidal oscillation is
already apparent in the jet interior.  In simulation P the emission
knots, visible in Figure 9 between $25< z/R_0 < 35$, have spacing
$\lambda \lesssim 4R_0$ and move with $v_w\sim 0.85a_e \sim 0.45u_j$
where $u_j \sim 1.9 a_e$.  It is interesting that the emission knots
appear in simulation P containing only a very weak toroidal field
component.  In simulation T the stronger toroidal field component might
have been expected to be more conducive to knot formation via a pinch
instability.  We will show that the presence of flow outside the jet in
simulation P has suppressed asymmetric structures and allowed the knots
to form.

\vspace{-0.7cm}
\section{Theoretical Analysis}

\vspace{-0.2cm}
\subsection{Stability Theory}

Let us model the jet as a cylinder of radius $R_j$, having a uniform density,
$ \rho _j$, a uniform internal axial magnetic field, B$_j$, and a
uniform velocity, u$_j$. The external medium is assumed to have a
uniform density, $ \rho _e$, a uniform external axial magnetic field,
B$_e$, and a uniform velocity, u$_e$. The jet is established in static
total pressure balance with the external medium where the total static
pressure is $ p_j^{*}\equiv p_j+B_j^2/8\pi =$
$p_e^{*}=p_e+B_e^2/8\pi $. Inclusion of a small non-uniform toroidal
magnetic field component like that used in simulation P will not
significantly modify results obtained from a linear analysis involving small fluctuations incorporating
only axial magnetic fields, although we expect significant effects
associated with the much stronger toroidal magnetic field used in
simulation T, e.g., Appl \& Camenzind (1992), Appl (1996), Rosen et
al.\ (1999). The general approach to analyzing the stability properties
of this system is to linearize the one-fluid MHD equations of
continuity and momentum along with an equation of state where the
density, velocity, pressure, and magnetic field are written as $\rho
=\rho _0+\rho _1$, ${\bf u} ={\bf u}_0+{\bf u}_{1\text{,}}$
$p=p_0+p_1$, ${\bf B}={\bf B}_0+{\bf B}_1$, and subscript 1 refers to a
perturbation to the equilibrium quantity. Axial magnetic fields in
cylindrical geometry have been investigated in several articles (Ray
1981; Ferrari, Trussoni, \& Zaninetti 1981; Fiedler \& Jones 1984; Bodo
et al.\ 1989) although in general it has been assumed that there is no
flow in the external medium. The relatively slow jet expansion in the
numerical simulations is not expected to significantly modify local
results based on a completely uniform external medium (Hardee 1984). 

In cylindrical geometry a random perturbation of $\rho _1$, ${\bf
u}_{1\text{,}}$ $p_1$, and ${\bf B}_1$ to an initial equilibrium state
$\rho _0$, ${\bf u}_{0\text{, }}$ $p_0$, and ${\bf B}_0$ can be
considered to consist of Fourier components of the form
\begin{equation}
\label{1}f_1(r,\phi ,z)=f_1(r)\exp [i(kz\pm n\phi -\omega t)] \text{ ,}
\end{equation}
where the flow is along the $z$-axis, and $r$ is in the radial direction
with the flow bounded by $r=R_j$. In cylindrical geometry $k$ is the
longitudinal wavenumber, $n$ is an integer azimuthal wavenumber, for
$n>0$ the wavefronts are at an angle to the flow direction, the angle
of the wavevector relative to the flow direction is $\theta =\tan
(n/kR_j)$, and $+n$ and $-n$ refer to wave propagation in the clockwise
and counterclockwise sense, respectively, when viewed outwards along
the flow direction. In equation (1) $n=$ 0, 1, 2, 3, 4, etc. correspond
to pinching, helical, elliptical, triangular, rectangular, etc. normal
mode distortions of the jet, respectively.  Propagation and growth or
damping of the Fourier components is described by the dispersion
relation
\begin{equation}
\label{2}\frac{\beta _j}{\chi _j}\frac{J_{\pm n}^{^{\prime }}(\beta _jR)}{
J_{\pm n}(\beta _jR)}=\frac{\beta _e}{\chi _e}\frac{H_{\pm n}^{(1)^{\prime
}}(\beta _eR)}{H_{\pm n}^{(1)}(\beta _eR)}\text{ ,}
\end{equation}
where $J_{\pm n}$ and $H_{\pm n}^{(1)}$ are Bessel and Hankel
functions, the primes denote derivatives of the Bessel and Hankel
functions with respect to their arguments, and we have dropped the
subscript on the jet radius, i.e., $ R\equiv R_j$. In equation (2)
$$
\beta _j=\left[ -k^2+\frac{(\omega -ku_j)^4}{(\omega
-ku_j)^2(a_j^2+V_{Aj}^2)-k^2V_{Aj}^2a_j^2}\right] ^{1/2}\text{, }\beta
_e=\left[ -k^2+\frac{(\omega -ku_e)^4}{(\omega
-ku_e)^2(a_e^2+V_{Ae}^2)-k^2V_{Ae}^2a_e^2}\right] ^{1/2}\text{ ,} 
$$
and 
$$
\chi _j=\rho _j[(\omega -ku_j)^2-k^2V_{Aj}^2]\text{ , }\chi _e=\rho
_e[(\omega -ku_e)^2-k^2V_{Ae}^2]\text{ ,} 
$$
where $u_{j,e}$,  $a_{j,e}$ and $V_{Aj,e}$ are the flow, sound and Alfv\'en
speeds in the appropriate medium.

In general, each normal mode, $n$, contains a single ``surface'' wave
and multiple ``body'' wave solutions that satisfy the dispersion
relation. The behavior of the solutions can be investigated
analytically and we extend previous analytical 3D results here to
include for the possibility of an external medium with axial magnetic
field and flow. 

\vspace{-0.4cm}
\subsection{Analytical Approximations}

The $n=0$ pinch mode ``fundamental'' (not a surface wave in this case) wave solution can be found from the dispersion relation in the low frequency limit $\omega \rightarrow 0$ and $k \rightarrow 0$.  In this limit $\beta_jR \rightarrow 0$,  $\beta_eR \rightarrow 0$ and the dispersion relation becomes
$$
\chi_j \approx -\chi_e(\beta_jR)^2[ln(\beta_eR)-i\pi/2]/2 \text{ ,}
$$
where we have used 
$[J_{0}^{^{\prime }}(\beta _jR)]/[
J_{0}(\beta _jR)] \rightarrow -\beta_jR/2$ 
and
$[H_{0}^{(1)}(\beta _eR)]/[H_{0}^{(1)^{\prime}}(\beta _eR)] \rightarrow
(\beta_eR)[ln(\beta_eR)-i\pi/2]$.  
As $(kR)^2 \rightarrow 0$ faster than
$ln(\beta_eR) \rightarrow -\infty$, the real part of the solution to the dispersion relation remains
nearly unmodified by the presence of a magnetized wind around the jet and is
given by (Paper I)
\begin{equation}
\label{3}\frac \omega k\approx u_j\pm \left\{ \frac 12\left( V_{Aj}^2+\frac{
V_{Aj}^2a_j^2}{a_{msj}^2}\right) \pm \frac 12\left[ \left( V_{Aj}^2+\frac{
V_{Aj}^2a_j^2}{a_{msj}^2}\right) ^2-4\frac{V_{Aj}^4a_j^2}{a_{msj}^2}
\right] ^{1/2}\right\}^{1/2}\text{ ,}
\end{equation}
where $a_{msj}\equiv \left( a_{j}^2+V_{Aj}^2\right) ^{1/2}$is the
magnetosonic speed. The imaginary part of the solution is vanishingly
small in the low frequency limit. These solutions are related to fast
($+$) and slow ($-$) magnetosonic waves propagating with ($u_j+$) and
against ($u_j-$) the jet flow speed $u_j$, but modified by the
jet-external medium interface. The unstable growing solution is
associated with the backwards moving (in the jet fluid reference frame)
slow magnetosonic wave.  The growth rate can only be determined by
numerical solution of the dispersion relation.

The $n>0$ higher order mode ``surface'' wave solutions can also be found from the dispersion relation in the low frequency limit. In this case the dispersion relation becomes
$$
\chi_j \approx -\chi_e \text{ ,}
$$
where we have used 
$[J_{\pm n}^{^{\prime }}(\beta _jR) H_{\pm n}^{(1)}(\beta _eR)]/[
J_{\pm n}(\beta _jR)H_{\pm n}^{(1)^{\prime}}(\beta _eR)] \rightarrow
-(\beta_eR)/(\beta_jR)$.  
The solution is modified by the presence of a magnetized wind around the jet and is given by
\begin{equation}
\label{4}\frac \omega k=\frac{u_e+\eta u_j}{1+\eta }\pm i\frac{\eta ^{1/2}}{
1+\eta }\left[ (u_j-u_e)^2-V_{As}^2\right]^{1/2}\text{ ,}
\end{equation}
where $\eta \equiv \rho _j/\rho _e$ and $V_{As}^2\equiv (\rho _j+\rho
_e)(B_j^2+B_e^2)/(4\pi \rho _j\rho _e)$. Growth corresponds to the plus
sign in equation (4), but these higher order surface modes are
predicted to be stable when $(u_j-u_e)^2-V_{As}^2<0$. When there is no
flow and no axial magnetic field outside the jet, equation (4) reduces
to equation (4a) in Paper I.  In this case the stability condition can be written as $u_j^2-(1+\eta)V_{Aj}^2<0$. These surface modes are strongly
influenced by flow and magnetic field in the medium outside the jet. In
the dense jet limit, i.e., $\eta \rightarrow \infty $ and
$V_{As}^2\rightarrow \eta V_{Aj}^2$, equation (4) becomes $\omega
/k\approx u_j\pm V_{Aj}$.  The present reanalysis in this limit reveals
an error in paper I where we stated incorrectly that $\omega /k\approx
u_j\pm V_{Aj}$ applied when the jet was sub-Alfv\'enic and stable
independent of the jet density. The unstable growing solution is
associated with the backwards moving (in the jet fluid reference frame)
Alfv\'en wave. When a jet is trans-magnetosonic and trans-Alfv\'enic
the growth (damping) rates of these wave modes can only be determined
by numerical solution of the dispersion relation.

The ``body'' wave solutions are somewhat modified by the presence of a
magnetized wind surrounding a jet and in the limit $\omega \rightarrow
0$ and $k \neq 0$ the dispersion relation becomes
$$
\beta_jR -n\pi/2 -\pi/4 \approx \pm\frac{\pi}{2}\left[1 - C_n\right] \text{ ,}
$$
where we have used $J_{n}(z) \approx (2/\pi z)^{1/2}cos(z-n\pi/2-\pi/4)$, $\theta = cos^{-1}\epsilon \approx \pi/2 - \epsilon$,
and $C_n << 1$ is a correction term given by
$$
C_n = \frac{2}{\pi}\frac{\chi _e}{\chi _j}\frac{\beta _j}{\beta _e}\frac{
H_n^{(1)}(\beta _eR)}{H_n^{(1)^{\prime }}(\beta _eR)}\left(\frac{\pi \beta_jR}{2}\right)^{1/2}J_{n}^{^{\prime }}(\beta _jR)\text{ .} 
$$
The solutions are given by
\begin{equation}
\label{5}kR\approx \frac{(n+2m+1/2)\pi/2-\pi/2(1- C_n)}{\left[
\{M_{msj}^2/[1-(M_{msj}/M_jM_{Aj})^2]\}-1\right] ^{1/2}} \text{ ,}
\end{equation}
where $m\geq 1$ is an integer When the external medium is unmagnetized
and there is no wind  $\chi _e|_{\omega = 0}=\rho
_e[u_e{}^2-V_{Ae}^2]k^2 = 0$ and $C_n = 0$.  Previously we found that
unstable body wave solutions exist only when the denominator in
equation (5) is real and that body mode growth rates, with the
exception of the 1st pinch body mode, remain small unless the jet is
sufficiently supermagnetosonic.  Thus, we do not expect to find
structure attributable to higher order, $n > 0$, body modes in the
present simulations.

\vspace{-0.5cm}
\subsection{Numerical Solutions}

We have investigated the effects of external magnetic fields and winds
on jet stability by solving the dispersion relation numerically using
root finding techniques.  The analytical expressions provide an initial
estimate for solutions in the low frequency limit.  We have considered
a set of parameters relevant to numerical simulations P \& T at a
distance $z \sim 18R_0$ from the inlet.  These parameters are given in
Table 2.  
\vspace{-0.75cm}
\begin{table}[h]
 \begin{center}
 \caption{Parameters at $z \sim 18R_0$ \label{tbl-2}}
 \vspace{0.1cm}
 \begin{tabular}{ c c c c c c c c c} \hline \hline
     {\bf Simulation} & $\eta$ & {$u_j/a_e^{ms}$} & {$u_e/a_{e}^{ms}$} &    {$M_{j}$} & {$M^{A}_{j}$}   & {$M^{ms}_{j}$} & $p_j/p^*_j$ & $p_e/p^*_e$  \\ \hline
  P & 0.3 & 2.20 & 0.77 & 2.23 & 1.43 & 1.20 & 0.33 & 0.80     \\
  T & 0.3 & 2.20 & 0.00 & 2.23 & 1.43 & 1.20 & 0.33 & 1.00   \\
 \hline
\end{tabular}
\end{center}
\end{table}
\vspace{-0.5cm}
Solutions to the dispersion relation can be sensitive to the
wind speed and to the presence or absence of a magnetic field in the
wind/external medium.  Solutions that are growing in one parameter
regime can be damped in another parameter regime.  Thus, we have
computed and show both growing and damped solutions to the dispersion
relation for an unmagnetized and a magnetized wind at wind speeds from
zero up to those observed in simulation P.  The root finding
techniques were not always capable of converging to a solution at all
frequencies or of finding a solution unless a very accurate initial
estimate could be supplied.  Solutions generally consist of a growing
and damped pair of roots and it was not always possible to find both
solutions; e.g., the root finder would not converge to the damped
fundamental pinch mode solution associated with the outwards moving (in the
jet fluid reference frame) slow magnetosonic wave.  Not all of the
possible solutions are shown.  For example, the fast magnetosonic wave
solutions associated with the fundamental pinch mode are not included and
various additional ``wind'' associated solutions found in the numerical
investigation are also not shown.  In general, these additional
solutions are purely real or damped, and will not affect the stability
properties of the jet.

\vspace{-0.5cm}
\subsubsection{The Fundamental Pinch and Asymmetric Surface Modes}

In the absence of a magnetic field in the external environment the
behavior of solutions to the dispersion relation is shown for a
representative sample of wind speeds in Figure 10.  Solutions
corresponding to the `backwards' moving waves (in the jet fluid frame)
are shown in the leftmost panels and solutions corresponding to the
`forwards' moving waves (in the jet fluid frame) are shown in the
rightmost panels. The wind speeds are indicated in the panels in units
of the magnetosonic speed in the external medium, in this case equal to
the sound speed in the external medium. The panels for an unmagnetized
external medium with no external wind are indicative of the expected
stability properties of the jet in simulation T.  In simulation T there
is little wind or magnetic field external to the jet, i.e., $u_e=0$ and
$p_e/p^*_e=1$.  The higher order ($n>1$) surface modes grow much faster
than the pinch fundamental mode and would be expected to dominate the
dynamics.  

\begin{figure}[h]
\vspace{12.9cm}
\includegraphics{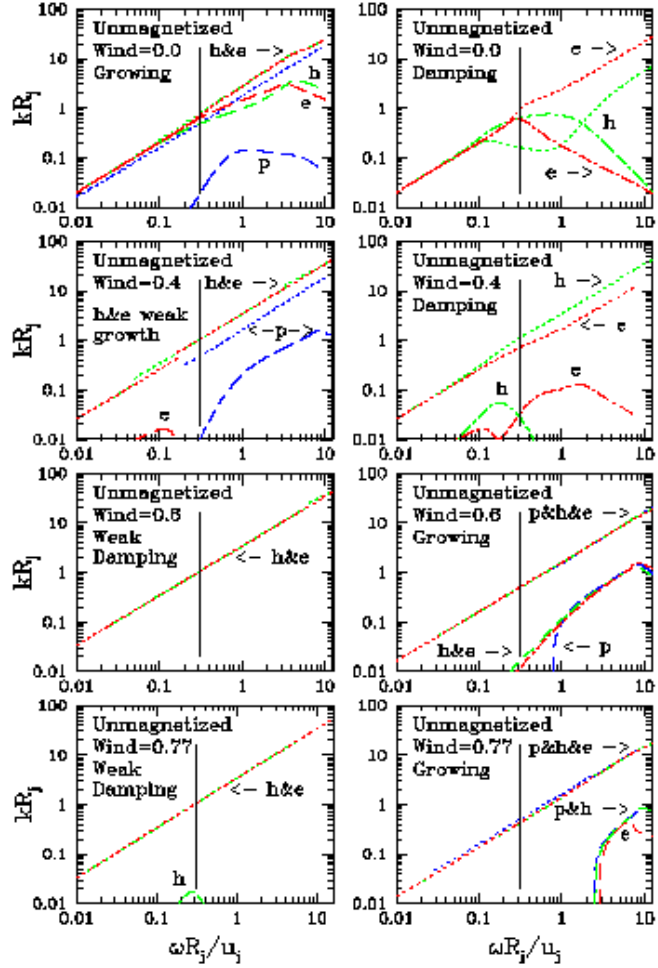}
\caption {\footnotesize \baselineskip 10pt  Solutions to the dispersion relation
for surface pinch (\Blue{p}), helical (\Green{h}), and elliptical
(\Red{e}) wave modes. Dotted lines indicate the real part of the
wavenumber.  Dashed (growing) and dash-dot (damping) lines indicate the
imaginary part of the wavenumber.  The vertical line indicates the
location of the precession frequency used in the simulations but
normalized to the jet radius and velocity at $z=18R_0$. In the panels
the wind speed is indicated in units of $a_e^{ms}$.}
\end{figure}

As the wind speed is increased the growth (damping) rates of helical
and higher order surface modes are, at first, greatly decreased as
predicted analytically by equation (4) at low frequencies (compare the
top two rows of panels in Figure 10). For these lower wind speeds
growth of the fundamental pinch mode is enhanced by the presence of an
external wind and can greatly exceed that of the helical and higher
order asymmetric modes. There is a relatively abrupt change in behavior
of the solutions when the wind speed is about 50\% of the sound speed
in the external medium.  At lower wind speeds the `backwards' moving
waves are growing and the `forwards' moving waves are damped. At higher
wind speeds the `forwards' moving helical and higher order surface
modes move with speed comparable to the fundamental pinch mode and are
growing, and the `backwards' moving waves are damped.  At these higher
wind speeds growth of the pinch and higher order modes are comparable
and relatively large, albeit at high frequencies only.   However, as
the wind speed increases further all growth rates diminish.  In
general, growth of the higher order modes is suppressed as the result
of an external wind.
   
The solutions to the dispersion relation are very sensitive to a
magnetized wind and this sensitivity to relatively low speed magnetized
winds is shown in Figure 11.  
\begin{figure}[h]
\vspace{13.0cm}
\includegraphics{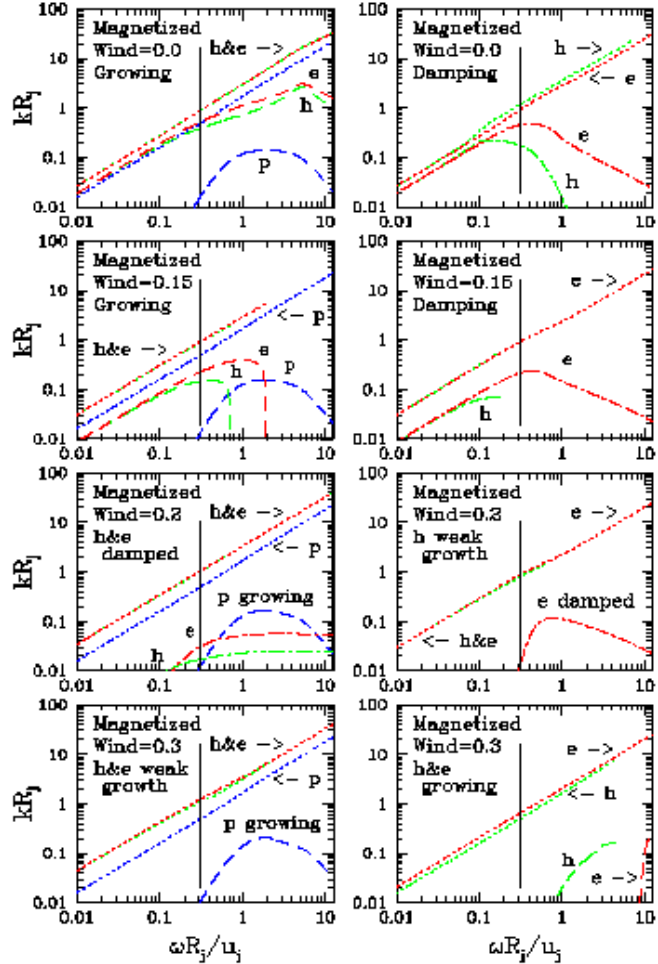}
\caption {\footnotesize \baselineskip 10pt  Surface mode solutions to the dispersion
relation for a magnetized external medium at lower wind speeds.
Notation is identical to that of Figure 10.}
\end{figure}
As in Figure 10 solutions corresponding
to the `backwards' and `forwards' moving waves (in the jet fluid frame)
are shown in the left and right columns, respectively.  Here the
external magnetic field has a value similar to that seen in simulation
P.  Note that the external magnetic field has a minimal effect on the
growth rates in the absence of a wind (compare the top left panel in
Figure 11 with the top left panel in Figure 10).  Comparison between
the leftmost panels in the top and second row in Figure 11 shows a
reduction in helical and elliptical growth rates at low frequencies
predicted analytically by equation (4).  In general, increase in the
wind speed stabilizes the higher order surface modes at the lower
frequencies while leaving the fundamental pinch mode growth rate
unchanged.  However, when the `forwards' moving helical and higher
order surface modes move with speed comparable to the fundamental pinch
mode they are growing and the maximum growth rate of the pinch and
higher order modes is again comparable (see the bottom panels in Figure
11), albeit growth of the higher order modes is restricted to high
frequencies. We note here that we do not confirm the result for the
wave speed of the higher order surface modes in the stable region as
was stated in paper I.  The wave speed is given properly by equation
(4) in this paper with the previously stated result, $v_w \approx u_j +
V_{Aj}$, only appropriate in the dense jet limit.  The error in paper I
resulted from an apparent but not true convergence of the root finding
routine.

There is a relatively abrupt break point in behavior of the solutions
when the magnetized wind speed is about 35\% of the magnetosonic speed
in the external medium, and solutions at higher magnetized wind speeds
are shown in Figure 12.  
\begin{figure}[h]
\vspace{9.7cm}
\includegraphics{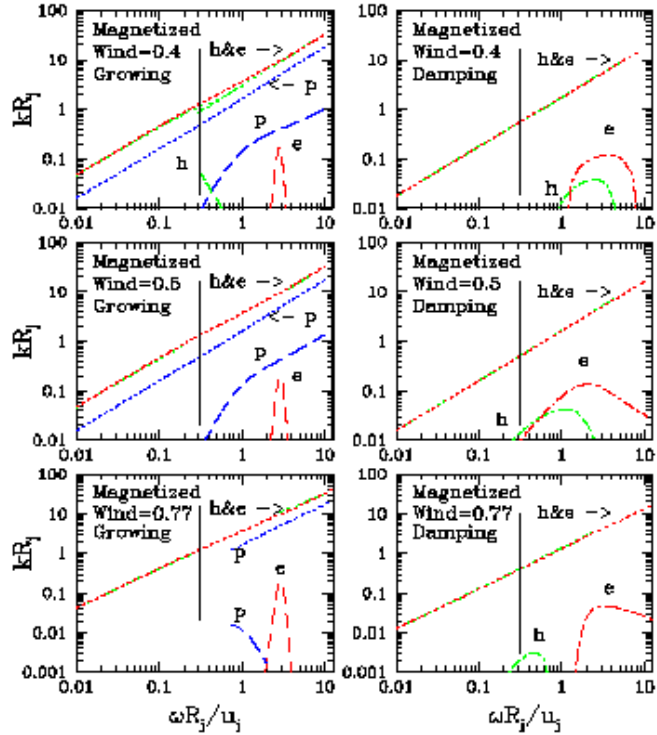}
\caption {\footnotesize \baselineskip 10pt  Surface mode solutions to the dispersion
relation for a magnetized external medium at higher wind speeds.
Notation is identical to that of Figures 10 \& 11.  Note the different
vertical scale in the bottom two panels (required to show the greatly
reduced growth and damping rates).}
\end{figure}
Once again the `backwards' and `forwards'
moving solutions are shown in the left and right columns,
respectively.  In general, higher order surface mode growth rates are
significantly reduced by these higher speed magnetized winds.  In these
magnetized wind cases our root finding techniques were unable to
converge to solutions at lower frequencies for the pinch fundamental mode
at the highest wind speed and in an intermediate frequency range where
we assume that the helical surface mode will exhibit a narrow
instability range similar to that exhibited by the elliptical surface
mode.  Nevertheless, with the exception of the highest wind speeds the
fundamental pinch mode is the most unstable and should dominate the
dynamics.  The computations in the bottom two panels indicate that the
magnetized wind observed in simulation P at the termination of the
simulation reduces the growth rate of the pinch mode by about an order
of magnitude, and restricts growth of the higher order surface modes to
a very narrow frequency range.  Clearly the fast magnetized wind in
simulation P is responsible for the stability of the jet to far beyond
the Alfv\'en point.

\vspace{-0.3cm}
\subsubsection{The Body Modes}

The analytical approximations at low frequency indicate that the real
part of the body mode solutions will be only minimally modified by the
presence of an external magnetic field or wind.  The extent to which a wind modifies the growth rates at higher frequencies is shown in Figure 13.
\begin{figure}[htb!]
\vspace{9.7cm}
\includegraphics{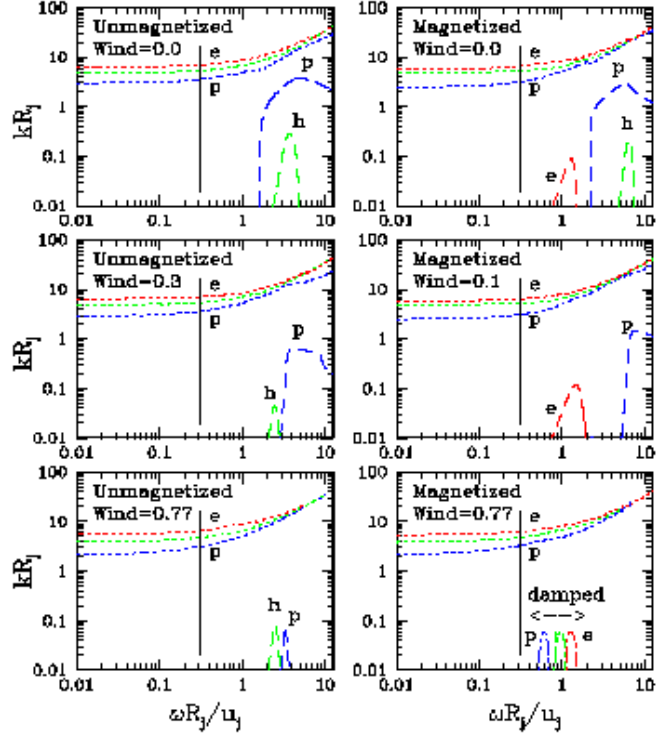}
\caption {\footnotesize \baselineskip 10pt  Pinch, helical and elliptical first body
mode solutions to the dispersion relation for an unmagnetized (left
column) and a magnetized (right column) external wind medium.  Notation is identical to that of Figures 10, 11, \& 12.}
\end{figure}
Even in the absence of a wind, the low magnetosonic Mach numbers
guarantee that body mode growth rates, with the exception of the first
pinch body mode, will be small.  The first pinch body mode has a
significant growth rate when the wind speed is low.  The addition of a
wind and, in particular, of a magnetized wind reduces these growth
rates even further and at the highest magnetized wind speed the body
modes are damped.

\vspace{-0.7cm}
\section{Interpretation of Simulation Results}

In simulations P and T the flow velocity is sub-Alfv\'enic for the
inlet conditions and thus enters the computational grid stable to
helical and higher order modes of jet distortion,  although not stable
to the fundamental pinch mode.  While in both simulations the jet
becomes super-Alfv\'enic within a few jet radii of the inlet, both jets
would remain stable beyond this point even in the absence of an
external magnetized wind as a result of their density.  For example, in
simulation P the jet should remain stable to helical and higher order
modes until $z \sim 10R_0$ when $u_j^2 - (1 + \eta) V_{Aj}^2 > 0$,
where $\eta \equiv \rho_j/\rho_e$ and $V_{Aj} \equiv [B_j^2/(4\pi \rho
_j)]^{1/2}$.  However, at this distance the wind speed has already
risen sufficiently, $u_e \sim 0.2 a_e$, so that $(u_j-u_e)^2 - (1 +
\eta) V_{Aj}^2 < 0$ and the jet is stable to helical and higher order
modes even in the absence of a magnetic field in the wind.  Simulation
P remains stable to helical and higher order modes out to $z \sim
30R_0$ where finally $(u_j-u_e)^2 - V_{As}^2 > 0$, where $V_{As} \equiv
[(\rho _j+\rho _e)(B_j^2+B_e^2)/(4\pi \rho _j\rho _e)]^{1/2}$.
Simulation T exhibits no such magnetized wind and destabilizes to
helical and higher order modes when $u_j^2 - (1 + \eta) V_{Aj}^2 > 0$
at $z \sim 12R_0$.

In general, normal mode solutions come in pairs  where the slightly
faster moving wave (forwards moving in the jet fluid frame) is damped
and the more slowly moving wave (backwards moving in the jet fluid
frame) is growing.  As the wind speed increases the spatial growth rate
of helical and higher order surface modes decreases as the velocity
shear decreases and the wave speed, measured in the stationary observer
frame, increases.  However, the fundamental pinch mode wave speed (in
the jet fluid frame a backwards moving slow magnetosonic wave) and
growth rate remain relatively unchanged as wind speed increases.  At
some intermediate wind speed, the faster moving (forwards moving in the
jet fluid frame) helical and higher order surface mode wave speed
becomes comparable to the speed of the fundamental pinch mode and now
these waves are growing whereas the slower moving helical and higher
order surface modes are damped.  Here the fundamental pinch and the
helical and higher order surface modes are growing with comparable
growth rates.  The intermediate wind speed value depends on the
magnetization of the wind and occurs at much lower wind speed for a
magnetized wind.  As wind speeds rise above this intermediate speed all
growth rates are reduced, but a larger wind speed is required to
achieve reduction in the fundamental pinch mode growth rate.

The different dependence of the fundamental pinch and the higher order
surface mode growth rates to the presence of an external wind explains
the  stabilization of simulation P to a distance well beyond the
Alfv\'en point, and also explains the presence of emission knots in the
jet in simulation P at intermediate simulation times.  In particular,
the emission knots, visible in Figure 9 at distances $25 < z/R_0 < 35$,
with $\lambda \lesssim 4R_0 \sim 3.5R_j$ and with $v_w\sim 0.85a_e \sim
0.45u_j$ can be identified with the fundamental pinch mode.   If we
apply the parameters given in Table 2 for simulation P we find that the
observed knot spacing corresponds theoretically to an angular frequency
$\omega R_j/u_j \sim 1$.  This frequency corresponds to the maximum
growth rate for an unmagnetized stationary external medium (see the top
left panel in Figure 10) and to the lowest frequency at which the
growth rate is comparable to the maximum growth rate in the presence of
a low speed magnetized external wind (see the bottom left panel in
Figure 11).  At these intermediate times and at this distance from the
inlet the large magnetized winds seen at the end of the simulation have
not yet developed.  The lower magnetized wind speeds at intermediate
times, $u_e \sim 0.15 - 0.35 a_e^{ms}$, suppress the helical and higher
order modes leaving the fundamental pinch mode to dominate and produce
the observed emission knots.  The absence of a significant wind in
simulation T results in no suppression of helical and higher order
surface mode instability.  Thus, in simulation T these normal modes
dominate the dynamics at all times and have suppressed the development
of the fundamental pinch mode and accompanying emission knots.

In order to make some quantitative comparison with structures observed
in the numerical simulations at large distances from the inlet we have
solved the dispersion relation for parameters typical to the jet at
$z/R_0 =$ 35 \& 50.  The parameters used for these distances are given
in Table 3 and $p_e/p^*_e=1$ and $a_e^{ms} = a_e$; recall that Table
2 contains jet parameters for $z/R_0=18$.
\vspace{-0.75cm}
\begin{table}[h!]
 \begin{center}
 \caption{Parameters at $z \sim$ 35 \& 50R$_0$ \label{tbl-3}}
 \vspace{0.1cm}
 \begin{tabular}{ c c c c c c c c c} \hline \hline
     {Sim @ z/R$_0$} & $\eta$ & {$u_j/a_e$} & {$a_j/a_e$} & $\omega R_j/u_j$
& {$M_{j}$} & {$M^{A}_{j}$} & {$M^{ms}_{j}$} & $p_j/p^*_j$ \\ \hline
  P @ 35 & 0.35 & 2.0 & 0.93 & 0.36 & 2.15 & 1.60 & 1.25 & 0.40    \\
  P @ 50 & 0.45 & 2.0 & 0.85 & 0.41 & 2.35 & 2.00 & 1.50 & 0.46  \\
  T @ 50 & 0.45 & 2.2 & 1.00 & 0.37 & 2.20 & 1.46 & 1.22 & 0.35  \\
 \hline
\end{tabular}
\end{center}
\end{table}
\vspace{-0.5cm}

\noindent
In simulation P at $z/R_0 < 30$ the short wavelength, $\lambda \lesssim
5R_0$, low amplitude oscillation in $p^{*}$ and $v_z$ is consistent
with the weak fundamental pinch mode instability that remains at the
highest magnetized wind speeds.  The jet destabilizes at about $z/R_0 =
30$ and at $30 < z/R_0 < 50$ exhibits a dominant oscillation in $v_x$
($v_r$) with $\lambda \lesssim 8R_0 \sim 5.3R_j$, where $R_j \sim
1.5R_0$ at $z/R_0 = 40$.  The observed wavelength is consistent with a
growing elliptical distortion at frequency $\omega R_j/u_j \sim 0.4$
for the relatively low wind speeds, $u_e/a_e^{ms} \lesssim 0.2$ found
at $z/R_0 \sim 40$, and an elliptical distortion is evident in the
cross sections in panels 30 and 35 in Figure 6. At $z > 55R_0$ dominant
out of phase transverse velocity oscillations indicative of helical
twisting with $\lambda \sim 10R_0 \sim 6R_j$ moving with $v_w\sim 0.75
a_e \sim 0.40u_j$ are consistent with precession induced helical
twisting for the observed low wind speeds of $u_e/a_e^{ms} \sim 0.2$.
Thus, the initial precession frequency is preserved through the
instability point.  Movement of the jet center off the $z$-axis appears
at $z \geq 40$ in the cross section panels in Figure 6.

In simulation T at $z < 40 R_0$ a region of growing amplitude oscillations in
$p^{*}$, $v_z$ and $v_x$ with $\lambda \lesssim 8R_0$
can be readily identified with an elliptical distortion seen in cross
sections in Figure 6 from $10 \leq z/R_0 \leq 30$.  We find that the
observed elliptical distortion is consistent with the elliptical
surface mode solutions determined from parameters appropriate to
simulation T at $z/R_0 = 18$ with $\lambda \sim 6.5R_j$ when $\omega
R_j/u_j \sim 0.25$.  At $z > 40R_0$ dominant out of phase transverse
velocity oscillations indicative of helical twisting with $\lambda \sim
10R_0$ moving with $v_w \sim 0.41u_j$ are readily identified with the
expected precession driven helical surface mode solution as was found
for simulation P.  The simulations indicate and the theoretical solution to
the dispersion relation confirms that the helical mode is relatively
insensitive to the exact values of the Mach number or to the wind
magnetization and low wind speed found in the simulations at these
large distances from the inlet.

In both simulations the magnitude of the toroidal velocity induced by
helical twisting declines approximately as $\exp [-2(r/R_j -1)]$ for
$r\geq R_j$, and the flow field remains well organized out to $r >
3R_j$ even though the magnitude of the toroidal velocity declines
rapidly.  The direction of the toroidal velocity field along the
line-of-sight oscillates with a sinusoidal wavelength equal to the
helical wavelength, i.e., $\sim 10R_0$.  This toroidal velocity field
orients the poloidal magnetic field in simulation P parallel and
anti-parallel to the line-of-sight with the banded structure shown in
Figure 8.  In simulation T a similar toroidal velocity field, albeit
with magnitude $\approx 2/3$ of that in simulation P, cannot reorient
the poloidal field component to overcome the orientation of the initial
toroidal field component.

\vspace{-0.7cm}
\section{Implications}

We have shown that jets can be stabilized to the KH helical and higher
order asymmetric normal modes provided the velocity shear, $\Delta u
\equiv u_j - u_e$, between the jet and the external medium is less than
a ``surface'' Alfv\'en speed, $V_{As} \equiv [(\rho _j+\rho
_e)(B_j^2+B_e^2)/(4\pi \rho _j\rho _e)]^{1/2}$.  Thus, the presence of
a magnetic field in the external medium and the presence of an
outflowing wind in the external medium are stabilizing.  At relatively
high magnetized wind speeds all normal modes are effectively
stabilized.  At more modest wind speeds or when the jet is
sub-Alfv\'enic the fundamental pinch mode remains unstable while the
helical and higher order modes can be partially or completely
stabilized. In this case the fundamental pinch mode dominates the
dynamics and emission knots can be produced in an initially steady
flow, albeit the simulations considered here contain an initial
precessional perturbation.  The initially stable poloidal magnetic
field simulation shows that the small amplitude precessional
perturbation is effectively communicated down the jet to the point
where the jet becomes unstable.  We note here that the wave speed in
the stable region is not as stated in paper I.  The wave speed is given
properly by equation (4) in this paper with the previously stated
result, $v_w \approx u_j + V_{Aj}$, only appropriate in the dense jet
limit.  The precessional perturbation couples to growing helical jet
distortions in the super-magnetosonic regime which are identical for
the poloidal and helical magnetic field simulations.  The helical twist
leads to a very well organized toroidal velocity field within
(see Figure 2) and outside of the jet.  The toroidal velocity field
orients any poloidal magnetic field inside and outside the jet along
the direction of the toroidal velocity.  We note that mixing in
simulation P has spread the poloidal field out to $r \lesssim 2R_j$, but
the velocity field remains well ordered out to $r \sim 3R_j$.  Thus, a
poloidal magnetic field in the medium surrounding a jet could be ordered by a
helically induced toroidal velocity field out to larger distance than seen in
simulation P. 

\vspace{-0.4cm}
\subsection{Protostellar Jets}

Recent work has suggested that protostellar jets may operate in the low
magnetosonic Mach number regime as low magnetosonic Mach numbers are
expected for magnetically launched jets (Fendt \& Camenzind 1996; Lery
\& Frank 2000).  However, the low magnetosonic Mach number regime is
typically KH unstable to helical and higher order normal modes.  In
such a situation, helical and other higher order asymmetric distortions
would be expected to dominate the dynamics and appearance of these
jets.  This fact has been demonstrated by somewhat higher Mach number
numerical simulations (Xu, Hardee, \& Stone 2000). While there is
evidence for helical and other asymmetric distortions in protostellar
jets, many of these jets contain emission knots along a relatively
straight jet, e.g., the knotty HH\,34 jet which lies within a weak CO
outflow (Reipurth et al.\ 1986; Chernin \& Masson 1995; Devine et
al.\ 1997).  A more complex example is provided by the HH\,111 jet
(Reipurth et al.\ 1997) which shows emission knots in an initially
relatively straight jet but with an apparent sinusoidal oscillation and
staggered bow shocks at larger distance.  Here the jet is surrounded by
a bubble within a larger molecular outflow (Nagar et al.\ 1997;
Cernicharo \& Reipurth 1996). The HH\,34 and HH\,111 jets are shown in
Figure 14.

\begin{figure}[h]
\vspace{5.2cm}
\includegraphics{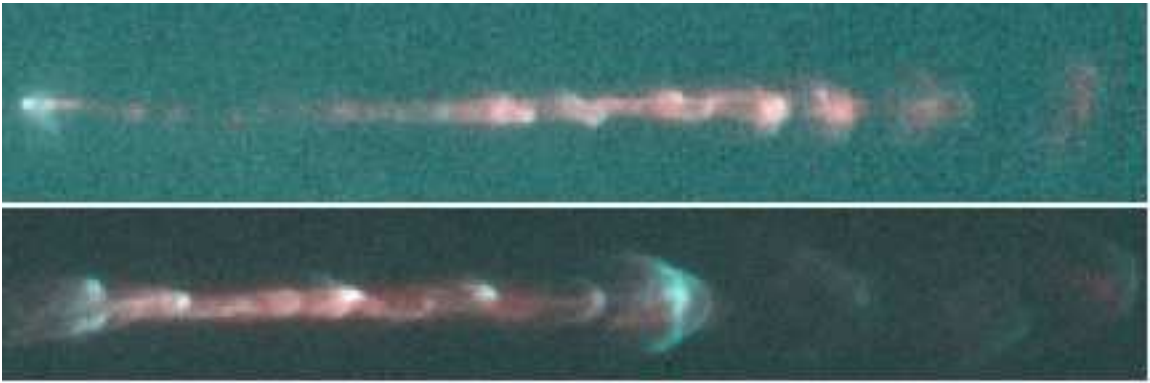}
\caption {\footnotesize \baselineskip 10pt  
Images of (top) HH\,34 and (bottom) HH\,111 (Devine et al.\ 1997; Reipurth et al.\ 1997).}
\end{figure}

In addition to HH\,34 and  HH\,111, observations have shown that other
protostellar jets are embedded in slower and less collimated molecular
outflows (see Richer et al.\ 2000 and references therein).  These
outflows typically surround but do not necessarily directly interact
dynamically with the more highly collimated jet, i.e., there is no
direct observational evidence for a significant wind in the material
immediately outside the jet.  That there might be a significant wind
environment around a more highly collimated jet is suggested by studies
of jet formation and collimation from an accretion disk, e.g., Meier,
Koide, \& Uchida (2001) and Begelman \& Blandford (1999) in the
extragalactic context and K\"onigl \& Pudritz (2000), Contopoulos \&
Sauty (2001) and Lery et al.\ (2002) in the protostellar context, or by
$X$-wind models in the protostellar context (e.g., Shu et al.\ 2000).
While the properties of the wind close to the jet are important, the
manner in which the wind is produced and its properties far from the
jet are not important.  The rapid exponential decline in velocity
fluctuations outside the jet implies an interaction layer of only a few
jet radii in thickness, and it does not matter whether a wind is
produced as part of the entrainment and mixing process (as in the
present simulations) or a wind originates in some other fashion (see
Lery et al.\ (2002) for some discussion of disk winds and $X$-winds,
and the relation between protostellar jets and molecular outflows).

Our present results suggest that if the magnetosonic Mach number of a
protostellar jet is sufficiently low, then a low speed magnetized wind
with 10\% of the jet speed (or significantly higher unmagnetized wind
speed) could stabilize the jet to helical and higher order modes of
asymmetric jet distortion while leaving the fundamental pinch mode to
grow and provide a trigger for knot spacing not too different from what
is seen in the protostellar jets.  Note that the present results
indicate that the wind need extend no more than a few jet radii
beyond the faster highly collimated jet material.  Spatial change in
the wind speed such as a lower wind speed at larger distance from the
source would allow helical and higher order modes to grow, triggered by
precession at the origin. Thus, it is possible that the interaction
between a low magnetosonic Mach number protostellar jet and an
appropriate outflowing wind could trigger knot formation with spacing
similar to that observed in the HH\,34 and HH\,111 jets without
requiring quasi-episodic jet injection on short timescales. 

The present numerical and theoretical results do not include the
effects of radiative cooling that are important on protostellar jets or
the effect of much higher jet density relative to the external medium.
In general, the inclusion of radiative cooling leads to greater
instability (Stone et al.\ 1997; Xu et al.\ 2000).  However, previous
2D (Hardee \& Stone 1997) and unpublished theoretical results  indicate
that the presence of significant magnetic field reduces the difference
between the stability properties of radiatively cooled and adiabatic
jets.

\vspace{-0.5cm}
\subsection{Extragalactic Jets}

The poloidally magnetized jet in simulation P shows surface
corrugations even before it destabilizes and after destabilization
shows significant cross section distortions in addition to elliptical
distortion and large scale helical twisting.  In simulation T the
presence of significant toroidal magnetic field component suppresses
the filamentation seen on the poloidally magnetized jet.  While the jet
in simulation T exhibits elliptical distortion and helical twisting all
higher order normal modes are suppressed.  This effect was observed in
paper I and in supermagnetosonic jet simulations containing a
significant toroidal magnetic field component (RHCJ).  No significant
mass entrainment occurs in the stable region in the poloidally
magnetized simulation P but significant mass entrainment accompanies
the development of elliptical distortion and helical twisting where the
jet destabilizes.  In simulation T, there is also evidence for mass
entrainment beginning where the jet destabilizes and elliptical
distortion appears.  Simulation P shows significantly less entrainment
and slowing of ordered flow than was observed in the comparable
``light'' jet simulation (simulation B in paper I), and this result
confirms that denser jets are more robust.  Simulation T could not be
run long enough to make a similar assessment, but it is clear from
paper I, the results here, and results from other work (RHCJ) that a
significant toroidal magnetic field helps to reduce entrainment and
helps to maintain jet collimation.

The line-of-sight appearance of the low magnetosonic Mach number
poloidally magnetized jets both here and in paper I when they become
unstable appears plume like.  Simulations at somewhat higher
magnetosonic Mach numbers with primarily poloidal or weak toroidal
magnetic field (Rosen, Hardee, Clarke, \& Johnson 1999; Rosen \& Hardee
2000) also appear plume like.  This set of poloidal and weak toroidal
magnetic field simulations seems most representative of the lower power
FR~I type extragalactic jets whose appearance has been argued (see
Bicknell 1994, 1995) to be the result of significant mass entrainment.
Surface turbulence from KH instability has been proposed previously as
a mechanism for the production of the high observed RM (rotation
measure) associated with some of the FR~I radio sources (Bicknell et
al.\ 1990), although other authors have pointed out quantitative
difficulties with this mechanism (Taylor \& Perley 1993; Ge \& Owen
1993).  Quantitative difficulties take the form of requiring an
excessively large magnetic field in a thin Faraday layer or require a
Faraday layer thicker than, for example, the dimension of the RM structure in
order to produce the observed RM with more modest magnetic fields.

The simulation results presented in RHCJ, Rosen \& Hardee (2000) and in
paper I revealed large scale organization in line-of-sight emission
image structure and in the accompanying B-field vectors associated with
helical twisting and mass entrainment.  These organized structures
extended throughout the mixing region and formed a relatively thick
layer around a central jet spine.  This occurred if polarization
B-field vectors were aligned with the jet flow or if polarization
B-field vectors were transverse to the jet flow if the toroidal
magnetic fields were sufficiently below equipartition.  The present
work reveals that the ordering is accomplished by the flow field
associated with helical twisting, should extend to at least a jet
diameter beyond the central jet spine, and can produce positive and
negative RM bands that alternate and lie across a jet. The sum total of
all the work suggests that ordering by the flow field outside the jet
spine requires a primarily poloidal magnetic field in the mixing region
and/or in the cocoon environment surrounding the jet/tail.  Our present
simulations do not specifically address the issue of poloidal field
alignment in the mixing region or in the external medium in
extragalactic sources but we note that the polarization vectors in the
3C\,31 jets suggest a poloidal orientation of the magnetic field in a
sheath even though the vectors suggest toroidal orientation of the
magnetic field in the jet spine (Laing 1996).

Recently Eilek \& Owen (2002) have considered the high RM structure
associated with radio source 3C\,465 in Abell cluster A2634 shown in Figure 15. 
\begin{figure}[hb!]
\vspace{10.4cm}
\includegraphics{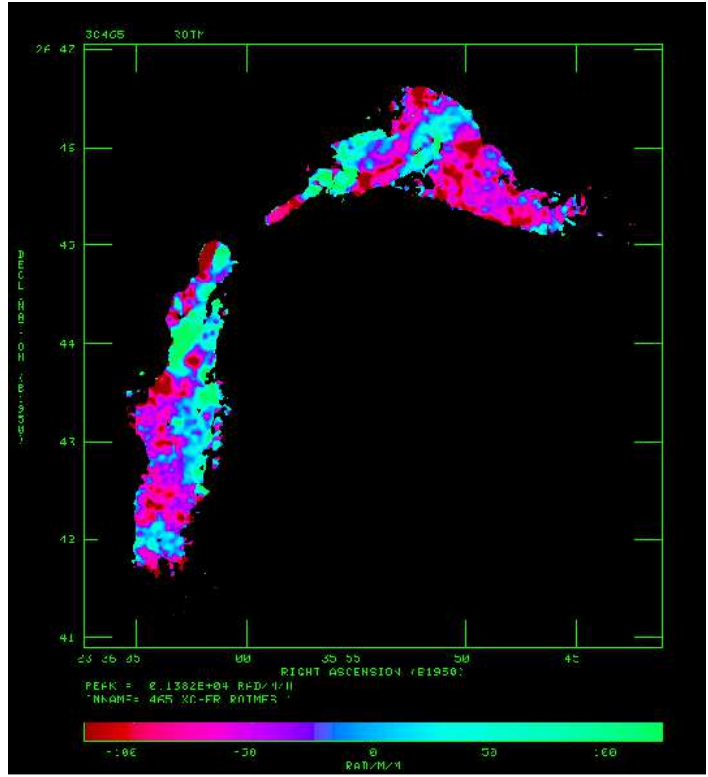}
\caption {\footnotesize \baselineskip 10pt  
Rotation measure distribution in Abell 2634 (3C~465).  The colors are chosen to distinguish positive (green \& light blue) and negative (red \& dark blue) values; $1 \arcmin$ is 35.2 kpc (from Eilek \& Owen 2002).}
\end{figure}
In particular the west jet/tail of this radio source, whose B-field
vectors more or less follow the tail, show positive and negative RM
bands that alternate and lie across the tail.  The emission and RM
images can be taken to suggest organization resulting from helical
twisting of the jet/tail flow.  The observations suggest a typical
scale size for the RM bands of $\sim 1 \arcsec$ or about 0.7~kpc, and
Eilek \& Owen find that a similarly dimensioned thin $\sim 0.7$~kpc
Faraday layer requires an excessively large, $\sim 96 \mu G$ and vastly
overpressured, $p_b/p_g \sim 420$ magnetic field to produce the
observed RM.  Thus, they conclude that the more likely candidate is a
thick Faraday screen associated with ordered much weaker cluster
magnetic fields.  However, in 3C\,465 the jet/tail diameter, as
inferred from the transverse extent of the radiating material, at the
location of the alternating RM bands is $\lesssim 5 \arcsec$ or about
3~kpc.  If we associate this distance with the thickness of a Faraday
screen then the magnetic field required to produce the observed RM is
reduced to $\sim 20 \mu G$ and $p_b/p_g \approx 25$.  If the jet/tail
is not in the plane of the sky, the line-of-sight through the sheath
will be longer.  While the required magnetic field is still high, we
conclude that RM structure in some radio sources may be associated with
large scale flow field organization of an ambient and/or jet magnetic
field in a thick sheath around a central jet spine.

\vspace {0.5cm}

P. Hardee and A. Rosen acknowledge support from the National Science
Foundation through grant AST-9802995 to the University of Alabama. The
numerical work utilized the Cray T90 at the San Diego Supercomputing
Center (operated under the auspices of the National Partnership for Advanced Computational Infrastructure, NPACI).

\end{document}